\documentclass[aps,prb,twocolumn,showpacs,superscriptaddress]{revtex4}
\usepackage{bm,amsmath,amssymb,latexsym,mathrsfs,graphicx,enumerate}
\usepackage[mathcal]{euscript}
\usepackage{hyperref}
\graphicspath{{figs/}}

\begin{document}
\date{\today : version 3.14}

\title{Statics of non uniform Josephson junction parallel arrays: model vs. experiment}
\author{F. Boussaha} 
\affiliation{LERMA, Observatoire de Paris, 77 avenue Denfert-Rochereau, 75014 Paris, France}
\author{J.G. Caputo} 
\affiliation{Laboratoire de Math\'ematiques, INSA de Rouen
   B.P. 8, 76131 Mont-Saint-Aignan cedex, France }
\author{L. Loukitch}
\affiliation{Laboratoire de Math\'ematiques, INSA de Rouen
   B.P. 8, 76131 Mont-Saint-Aignan cedex, France }
\author{M. Salez} 
\affiliation{LERMA, Observatoire de Paris, 77 avenue Denfert-Rochereau, 75014 Paris, France}

\begin{abstract}
We study experimentally and numerically the zero-voltage supercurrent vs. 
magnetic field of non-uniform arrays of Josephson junctions 
parallel-connected by a superconducting stripline. The measured curves are 
complex, unique and in excellent agreement with numerical simulations using 
a specially developed model. Using this, we can optimize the arrays 
to have any 
desired interference pattern. Such new devices could find applications in
magnetometry, quasiparticle mixers and detectors, flux-flow oscillators and
superconducting electronics.
\end{abstract}

\pacs{74.81.Fa, 85.25.Dq}

\maketitle
Josephson transmission lines have been experimentally and 
numerically investigated for their unique properties deriving from the 
Josephson non-linearity \cite{Barone,Likharev,u}. Such systems include
the "long" Josephson 
junction $(LJJ)$ and the parallel array of "small" superconducting tunnel 
junctions connected by a superconducting transmission line, 
called discrete Josephson transmission line $(DJTL)$ 
\cite{pk,nms,mghg,zbod}. "Small" vs. "long" qualifies the junction size 
relative to the Josephson length\cite{Likharev}
\begin{equation}\label{lambda1}
\lambda_J=\sqrt{\Phi_0/(\mu_0 J_c d)}~ ,
\end{equation}
where $\Phi_0=\hbar/2e$ is the reduced magnetic flux quantum, $J_c$ is 
the critical 
current density and d is the magnetic thickness\footnote{\label{1}The magnetic 
depth in a Josephson transmission line is given by $d=\mu_{\rm eff} t_d$ where 
$\mu_{\rm eff} = 1+(\Lambda_L/t_d)[\coth(t_1/\Lambda_L)
+ \coth(t_2/\Lambda_L)]$ where $\Lambda_L$  is the
London depth in niobium and td, t1 and t2 are respectively the thicknesses
of the dielectric layer (junction's tunnel barrier or stripline
dielectric) and of the lower and upper electrodes.}
between the junction 
electrodes. A $DJTL$ can also be viewed as a parallel array of $N-1$ 
$SQUID$ loops\footnote{\label{2}Superconducting parallel arrays of tunnel 
junctions bear
various names in the literature : "discrete Josephson transmission line"
($DJTL$) in the Josephson physics community \cite{u}, "parallel-connected
tunnel junction" ($PCTJ$) among $SIS$ mixer engineers \cite{sniis},
"superconducting quantum interference grating" ($SQUIGs$) \cite{mghg} or
"superconducting quantum interference filters" ($SQIFs$) \cite{ohs} in
magnetometry. In the limit $a\rightarrow 0$, "long Josephson junction" 
($LJJ$) 
and "distributed quasiparticle $SIS$ mixer" \cite{tbbsl95} also describe 
the same object from different perspectives.} \cite{nms,mghg,zbod}.
For magnetometry, such circuits have been considered recently
because they are more sensitive than $SQUIDs$, the response of their total 
zero-voltage supercurrent $I_{\rm max}$ to a magnetic field $H$ 
contains much narrower bumps. $I_{\rm max}(H)$ is then analog to 
the diffraction pattern of N-slit optical gratings. 
Recently non-uniform junction distributions have 
been considered to synthesize an $I_{\rm max}(H)$ curve with a 
single narrow peak at $H=0$, allowing an absolute magnetic zero reference 
as well as a higher sensitivity\cite{zr}. 
For $DJTLs$ where the Josephson 
coupling is spatially discontinuous and exists only in the junctions, the
difference $\varphi(x,t)$ between the superconducting phases of the top
and bottom electrodes obeys an inhomogeneous sine-Gordon 
equation\cite{cfv95}. When the phases vary little from one junction to
another, static and dynamic fluxons, i.e. $2\pi$ kinks of $\varphi$ 
can occur in $DJTLs$. 
\begin{figure}
\includegraphics[width=\linewidth,angle=0]{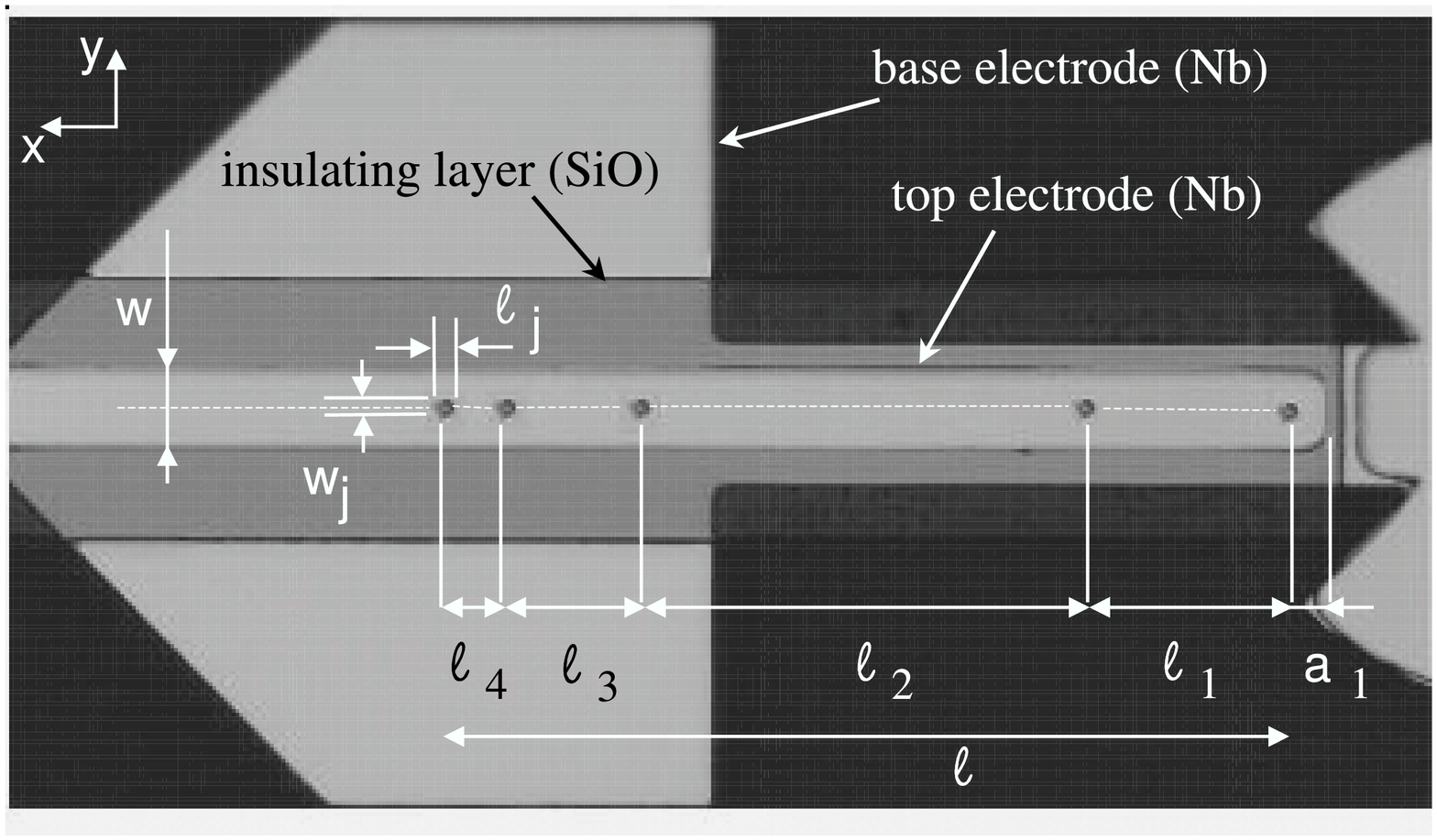}
\caption{Photograph of circuit "A" (see Table 1). The $1 \mu m^2$
junctions are made out of a $Nb/Al/AlOx/Nb$ trilayer, patterned using a
self-aligned process, where the upper Nb layer is removed by $RIE$ around
the junction mesas and replaced by $SiO$. A $Nb$ counter-electrode is then
sputtered to contact junctions and make the $Nb/SiO/Nb$ $rf$ tuning
microstrip circuit. This one, designed for broadband submillimeter-wave
$SIS$ mixing, is also a non-uniform $DJTL$ or $SQUIG$.}
\label{f1}
\end{figure}
In practice, arrays of small junctions are more flexible than long junctions.
They allow better impedance matching, their quasi-particle and surface dampings
can be adjusted separately and as we will see here their interference patterns
can be designed to suit specific needs.
For microwave detection, small junctions have been successfully used 
to make quantum-noise limited $SIS$ heterodyne mixers 
and detectors from $~100 GHz$ to $1000 GHz$, primarily for radioastronomy 
\cite{zr,sa}. These devices use photo-assisted quasiparticle tunneling 
(the damping term in the sine-Gordon equation).
Recently $SIS$ mixer designs have mingled with 
$DJTLs$\cite{sniis,tbbsl95}. In particular, it was demonstrated 
that a uniform $DJTL$ circumvents the fundamental bandwidth limitation of 
resistive mixers\cite{sa2}. We extended this concept to non-uniform arrays 
with various junction sizes and spacings, to optimize the impedance $Z(f)$ over 
virtually any desired submillimeter-wave band (below the superconductor's gap 
frequency). Such junction arrays behave as multipole bandpass filters\cite{sa2}. 
They have also been recently proposed as submillimetric direct detectors
In these applications, the Josephson effect brings excess noise and instabilities.
Therefore in practice, one quenches $I_{\rm max}$ by applying 
in the plane of the oxide layer a 
magnetic field corresponding to $\Phi=n\Phi_0$, where usually $n = 1$ or $2$. Yet 
the Josephson response to $H$ of complex, non-uniform $DJTLs$ is not straightforward, 
due to unknown flux internal distributions, possible phase slippage, ac modes. This 
initiated our study of Josephson phenomena in $SIS$-array mixers. This Letter focuses 
on the static behavior.

The non-uniform $SIS$-array mixers investigated (see Table 1 and Fig.~\ref{f1}) 
consist of $N=5$ $Nb/AlOx/Nb$ junctions embedded in a $Nb/SiO/Nb$ stripline of width 
$w= 5-7 \mu m$\cite{tn}. The array length is $l =80-86 \mu m$. Junctions 
have equal width 
and length (${\rm w_j} = {\rm l_j} \approx 1 \mu m$) and their spacings 
were optimized so that 
$Z (f) \sim  40 \Omega $ over $\sim 400-600 GHz$, for best $rf$ coupling in 
that range\cite{b06}. We successfully applied that design approach to $SIS$ mixers for 
a spaceborne instrument\cite{sniis}. The devices were made with a self-aligned, 
Selective Niobium Etch Process ($SNEP$) using optical lithography on fused quartz. 
Critical current densities $J_c$ range from $5$ to $13 kA cm^{-2}$, junctions have 
$V_g = 2.85 mV$ and subgap-to-normal resistance ratios of $15-20$. 
The Swihart velocity was measured in $LJJs$ located on the same wafer, to derive 
the specific junction capacitance for each batch. $C_s = 77$ (resp. $81$) 
$\pm 3 fF \mu m^{-2}$ for $J_c = 5$ (resp. $10$) $kA cm^{-2}$ are close to 
the assumed value $80 fF/\mu m^2$. With a $250-nm$ $SiO$ layer, the Swihart 
velocity $\bar{c}$ in the arrays is $\sim0.14 c$. 
In the microwave problem, junctions are point-like and arrays are 1D, since 
${\rm l_j} < w << \lambda = \bar{c}/f$ over the whole band. The "small junction" 
approximation is justified because ${\rm l_j} < \lambda_j \approx 5 \mu m$ 
In a $DJTL$ the size of fluxons at rest is defined 
by the discreteness parameter\cite{zbod} $\Lambda_J \equiv \lambda_J / a$ 
where $a$ is the junction spacing. 
$\Lambda_J>1$ implies that the flux permeates the array over 
several cells and that fluxons may move freely. For $\Lambda_J<1$, 
fluxons must overcome an energy barrier to move from each cell to its neighbor. 
In our case, there is no periodicity $a$ so our discreteness parameter is
$\Lambda_J^* \equiv \lambda_J/ l_{\rm av}$ where $l_{\rm av} = l/(N-1)$ 
is the $average$ distance between junctions. 
The design and 
performance of these non-uniform array mixers are detailed elsewhere\cite{sa2}.
\centerline{{\bf Table 1. Non-uniform $DJTL$ geometry}}
\centerline{(Dimensions of the junctions in $\mu m$, $w_j=l_j=1$).}
\centerline{$\begin{array}{|c|c|c|c|c|c|c|c|}\hline
{\rm Device} & N & w & a_1 & l_1 & l_2 & l_3 & l_4 \\ \hline
A       & 5 & 5 &   2   &   20  &   42  &   12  & 6 \\ \hline
B       & 5 & 7 &   2   &   18  &   44  &   11  & 7 \\ \hline
F       & 2 & 5 &   2   &   12  &   -   &   -   & - \\ \hline
\end{array}$}
We model the device shown in Fig.~\ref{f1}.
The phase difference between the top and bottom superconducting 
layers satisfies in the static regime
the following semilinear elliptic partial differential equation 
\cite{cfv95}
\begin{equation}\label{2dsg}
-\Delta \varphi + \frac{g(x;y)}{\lambda_J^2}\sin \varphi = 0,
\end{equation}
where $g(x;y)$ is $1$ in the Josephson junctions and $0$ outside and
we assume the same magnetic thickness $d$ in all the device. This
formulation guarantees the continuity of the normal gradient of $\varphi$,
the electrical current on the junction interface.

The boundary conditions representing an external current input $I$
and an magnetic field $H$ applied along $y$ are
\centerline{$\begin{array}{l}
\varphi_y|_{^0_w}=\left[ \mp \mu_0\nu  \frac{I}{2l} \right] d/\Phi_0
= \mp \nu \tilde{I}/(2l)\;,\\
\varphi_x|_{^0_l}=\left[H \mp \mu_0 (1-\nu)\frac{I}{2w} \right]d/\Phi_0
=\tilde{H} \mp (1-\nu) \frac{\tilde{I}}{2w}\;,
\end{array}$}
where $0\leq \nu \leq 1$ gives the type of current feed. When
$\nu = 1$ the current is only applied to the long boundaries 
$y=0,w$ (overlap feed) while $\nu=0$ corresponds to the inline feed. 
The
quantities $\tilde{H}$ and $\tilde{I}$ are respectively the normalized
magnetic field and current.

We consider long and narrow strips containing a few small junctions
of area ${\rm l_j} \times {\rm w_j}$ placed on the line $y=w/2$ and 
centered on 
$x=a_j$, $j=1,n$. For narrow strips $w<\pi$, only the first Fourier's mode 
needs to be taken into account\cite{cfgv96},then we write $\varphi$ as
\centerline{$\varphi(x,y) \approx  \nu \tilde{I}(y-w/2)^2/(lw) + 
\phi(x)~$.}
The first term of the equation takes care of the boundary conditions.
Thus, we obtain the following equation for $\phi$:
\centerline{$-\phi_{xx} + G(x)[w_j/(w\lambda_J^2)]\sin \phi = \nu\gamma/l\;,$}
where $\gamma=\nu \tilde{I}/w$.
We restrict ourselves to small junctions and 
we consider that the phase does not vary in the junctions. To
keep a continuous model, we decrease the size of the junction 
without neglecting the current crossing it. Then $g$ tends to a sum 
of delta functions\cite{cl05,cl06}:
\begin{eqnarray}
-\phi_{xx}+\sum_{i=1}^N d_j \delta(x-a_i)\sin\phi = \nu\frac{\gamma}{l},&&
d_j = \frac{w_j l_j}{w \lambda_J^2},\nonumber \\[-1.5ex]
\label{pb1D}\\[-1.5ex]
\phi_x|_{^0_l}  = \tilde{H} \mp\frac{1-\nu}{2w}\tilde{I}\;.&&\nonumber
\end{eqnarray}
Despite its theoretical aspect, the delta function allows
a detailed mathematical analysis of the solutions\cite{cl06}.

As opposed to standard models of this type of device
\cite{Barone,Likharev,mghg}, our approach does not neglect the variation of
$\varphi$ in the linear part of the circuit. This allows, for example,
to show the differences between inline and overlap current feed. The main
advantage of eq.(\ref{pb1D}) is that we can choose the
position $a_j$ and size $(w_j,l_j)$ of each junction and that we satisfy 
the matching conditions that exist in the original problem (\ref{2dsg}).
We have shown\cite{cl06} the periodicity of $\gamma_{\rm max}$
curve for an array where the junction positions are rational and analyzed
the differences between inline and overlap current feed, the effect of 
one remote junction on the array etc... There 
we established that for a device with sufficiently small junctions, 
the $\gamma_{\rm max}$ curve of eq.(\ref{pb1D}) tends to a simple function called 
the {\it magnetic approximation}.

For the device $A$ (resp. $B$) (see Table 1), 
$d_j\approx 0.025$. (resp. $d_j\approx 0.0178\dots$). For such small 
coefficients, the magnetic current which flows the junctions does not affect much
the phase in the linear parts of eq.(\ref{pb1D}). In this case,
the phase gradient is proportional to the magnetic field $\tilde{H}$ (see 
eq.(\ref{pb1D})).
We therefore assume that $\phi(x) \equiv \tilde{H}x + c$, 
so $\gamma=\sum_j d_j \sin(\tilde{H}a_j+c)$.
To find the $\gamma_{\rm max}(\tilde{H})$ curve for the magnetic approximation, 
we take the derivative of $\gamma$ with respect to $c$.
The values of $c$ such that $\partial \gamma / \partial c=0$ are
\centerline{$c_{max}(\tilde{H}) = \arctan
\left[\frac{\sum_{j=1}^N d_j \cos(\tilde{H}a_j)}
{\sum_{j=1}^N d_j \sin(\tilde{H}a_j)}\right]\;$, we obtain}
\begin{equation}\label{magn.app.}
\gamma_{max}(\tilde{H}) = \left|\sum_{j=1}^N d_j
\sin\left(\tilde{H} a_j + c_{max}(\tilde{H})\right)\right|\;.
\end{equation}
The absolute value is to guarantee that this is a maximum
and not an extremum.
This generalizes the standard approach\cite{Likharev,mghg}.
\begin{figure}
\includegraphics[height=\linewidth,width = 4cm ,angle=270]{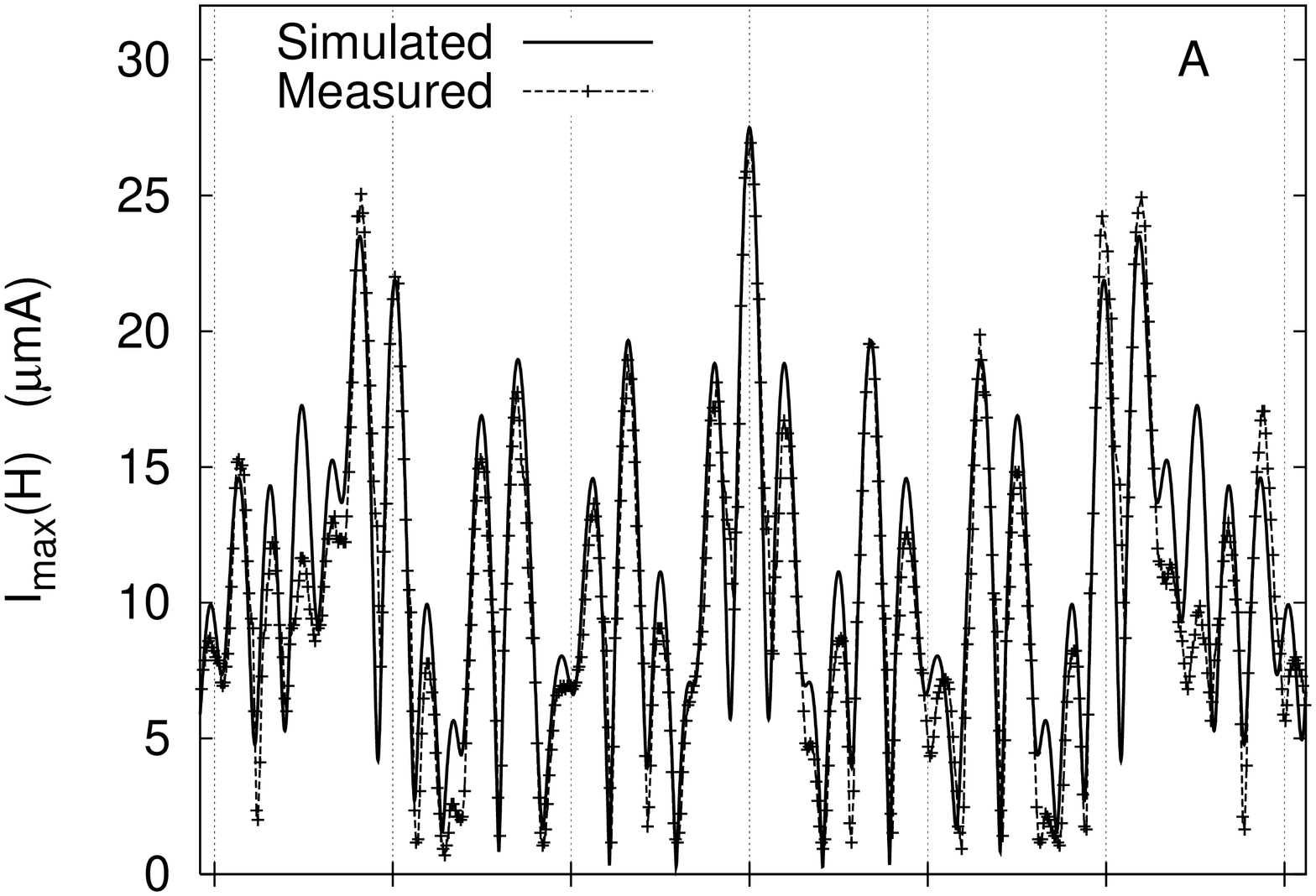}
\includegraphics[height=\linewidth,width = 4cm ,angle=270]{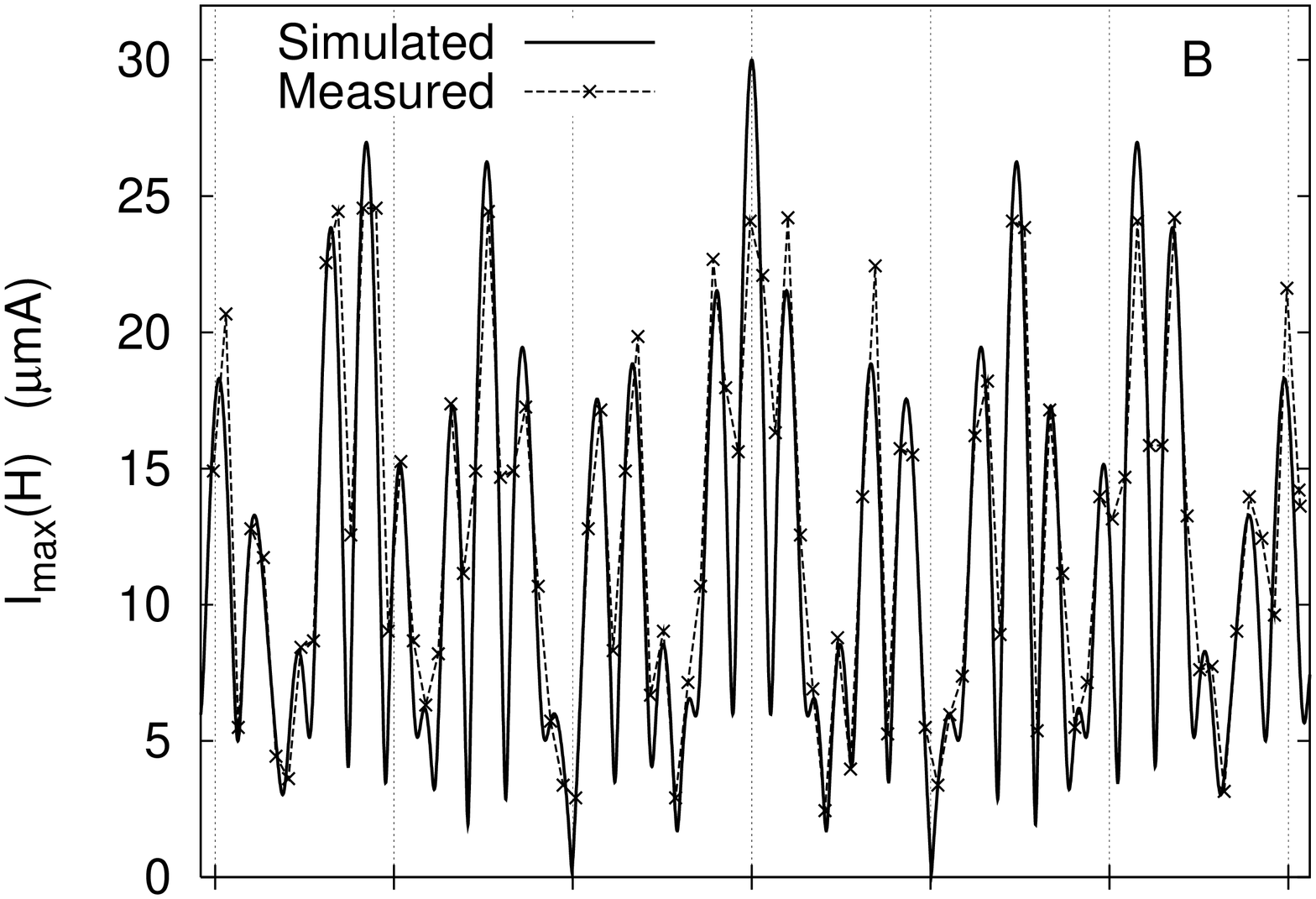}
\includegraphics[height=\linewidth,width = 4cm ,angle=270]{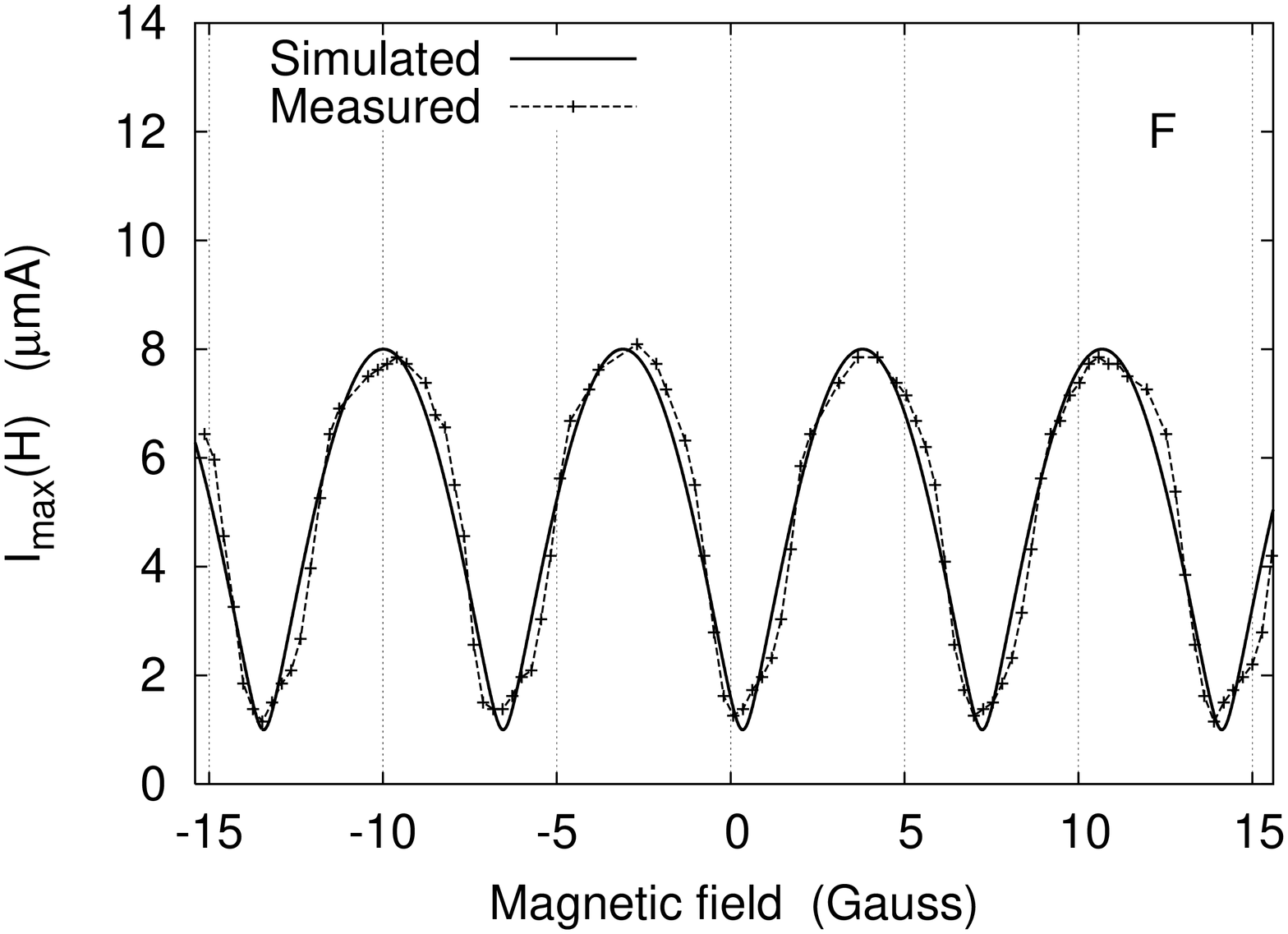}
\caption{Measured and simulated (see eq.(\ref{magn.app.})) $I_{\rm max}(\tilde{H})$in
the $V = 0$ state versus $H$ at $4.2 K$ for the arrays of junctions:
$A$, $B$ and $F$. Notice that for the device $F$, we simulate the device with
one junction with an area $\approx 1.1 \mu m^2$ and the second
$\approx 0.9 \mu m^2$.}
\label{f2}
\end{figure}
Fig.\ref{f2} shows the measured and simulated supercurrent $I_{\rm max}(\tilde{H})$ 
for the circuits $A$, $B$ and $C$, with the $H$ field applied along the y-axis. 
All circuits are from the same batch and $F$ was used to calibrate the measurements 
and validate the numerical model. As an additional test of the model, we computed 
$I_{\rm max}(\tilde{H})$ curves of arrays with $N=5$, $10$ and $20$ periodically 
spaced junctions, 
and obtained the same patterns reported in other works\cite{ks}. The experimental 
$I_{\rm max}(\tilde{H})$ curves were all measured at $4.2 K$. The magnetic field 
was produced by a superconducting electromagnet calibrated with a cryogenic 
Hall probe. Fig.\ref{f2} 
is a blow-up of the region near $H = 0$ where the features of the
array geometry become visible. On a larger $H$ 
scale ($300$ Gauss corresponds to $\Phi=\Phi_0$ in a $1\mu m^2$ junction) all curves 
follow the well-known Fraunhofer diffraction pattern given by the length $l_j$ of the
junctions. Within this envelope, the 
$I_{\rm max}(\tilde{H})$ curves of $A$ and $B$ follow complex yet 
reproducible patterns which are geometry-dependent, yet qualitatively similar.
The $I_{\rm max}(\tilde{H})$ curves 
of $A$ and $B$ are undisputedly different, showing the extreme sensitivity of the 
Josephson "interference" to junction distribution. 
In Figs.\ref{f2}A and B, the measured $I_{\rm max}(\tilde{H})$ curves were smaller
than expected from $I_c$. This can be due to the difference in magnetic 
thicknesses $d$
in the junction and microstrip that we neglect\cite{cfv95}.
The simulated curves 
of $A$, $B$ and $F$ are 
homothetically adjusted on the current axis by respectively the factors 
$260$, $270$ and $201$. For the $H$ 
axis, the needed adjustment factors are $0.8$, $0.8$ and $1$ and in the case 
of $F$ an offset of 2.8 Gauss is compensated for. 
For the SQUID (F) we had to slightly correct the area of one of the junctions
to fit $I_{\rm max}(\tilde{H})$ and get its smooth behavior near the minima 
instead of the
sharp behavior obtained for identical junctions\cite{cl06}.
The periodicity of $I_{\rm max}(\tilde{H})$ in the $SQUID$ is $H_0 = 4.8$ Gauss, 
close to the theoretical value $H_0 = 4.6$ Gauss, derived from a calculated 
effective magnetic permeability\ref{1} $\sim 1.68$ in the 
$Nb/SiO/Nb$ stripline. The ratio of the bumps average 
widths in the $N=5$ and $N=2$ cases is $\sim 8$, which is the ratio of the 
circuits lengths.  
For $l = 80 \mu m$, $\tilde{H}\approx  0.6$ Gauss corresponds to $\Phi_0$. The 
narrow ($0.5-0.8$ 
Gauss) bumps for the $N=5$ arrays therefore correspond to supercurrent vortices 
extending over the whole circuit, 
indicating that the phase distributes itself over the
whole device.
\begin{figure}
\includegraphics[height=\linewidth,width = 4cm,angle=270]{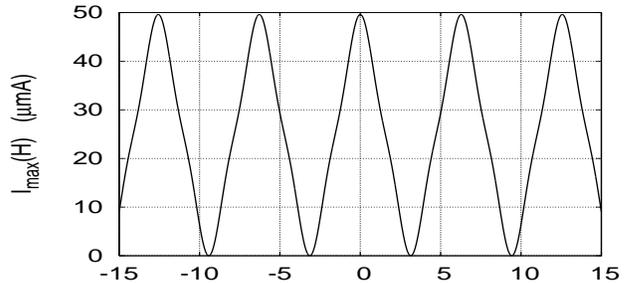}
\caption{Simulated $I_{\rm max}(\tilde{H})$ curve for an array of $5$ junctions such
(all lengths are given in $\mu m$) $w=5$, $a_1 = 2$, $l_1=21.2$, $l_2=10.6$,
$l_3=10.6$, $l_4=21.2$ and respective area of the junction are
(given in $\mu m^2$): $0.24$, $2.24$, $4.96$, $2.24$, $0.24$ and
$J_c=5 kAcm^{-2}$.}
\label{f3}
\end{figure}

Most experimental and numerical studies of 
fluxons in $DJTLs$ have dealt with uniform arrays\cite{pk,zows,dwtozs,tbbsl95} and 
large junction numbers ($N>10$). In such a $DJTLs$ with $\Lambda_J <1$, flux
quanta experience pinning and therefore the "nuls" in $I_{\rm max}(\tilde{H})$
do not reach $I=0$
but a $I_{\rm min} \neq 0$ floor which rises with $\Lambda_J^{-1}$ \cite{zbod}.
The non-uniform arrays reported here have different non-zero $I_{\rm min}$ values
for all minima and since all bumps in $I_{\rm max}(\tilde{H})$ are
clearly seen, fluxon pinning is moderate.
We expect mobile fluxons in the circuit $A$ because $\Lambda_J^*\approx 1.78 > 1$
($J_c = 5 kA cm^{-2}$).
We observed\cite{sa2} several constant-voltage branches in the $IVCs$ of $A$, 
$B$ and other non-uniform arrays, due to Josephson resonances. Their 
frequencies $f_n$ and their shapes suggest traveling fluxons. We have also 
seen such branches in the numerical solutions for given voltages.

Fluxon propagation in $LJJs$ has been extensively studied for
submillimeter-wave flux-flow oscillators ($FFOs$) in superconducting integrated 
heterodyne receivers\cite{u,ohs}. Non-uniform $DJTLs$ hosting fluxons 
offer advantages over $LJJs$ for $FFOs$\cite{tn}.
Whereas broadband 
$rf$ coupling is a challenge for the very-low impedance $LJJs$, non-uniform 
$DJTLs$ such as discussed in this Letter were optimally designed for this. 
Also, tunneling and stripline 
regions distinct in a $DJTL$ allow to adjust separately 
quasiparticle loss and surface loss parameters which 
jointly determine the oscillator mode (resonant vs. flux-flow)\cite{ks}. 
Such non-uniform $DJTLs$ could also be useful in fluxon-based electronics.

Our mathematical model, validated by measures, allows to develop non-uniform 
$DJTLs$ for magnetometry\ref{2} ($SQIFs$). It can also be used to improve 
their performance
as Josephson magnetic filters.  For instance, microwave circuits and sensors 
using quasiparticles 
should be most insensitive to $H$ noise. To reduce Josephson noise 
in $SIS$ mixers one biases them with a magnet on the zeroes of their 
$I_{\rm max}(\tilde{H})$ curve. 
But mixer stability is limited by how narrow the dips are at one zero. A specially 
designed non-uniform $SIS$ array with zeroes considerably flattened out and set at 
specified $H$ values would solve this problem. Just as an active antenna array 
can synthesize a given microwave beam shape, non-uniform $DJTLs$ can produce 
a given $I_{\rm max}(\tilde{H})$ curve and our model can be reversely applied to derive 
the required geometry. As an example, 
Fig.$\ref{f3}$ shows the simulated triangular-periodic $I_{\rm max}(\tilde{H})$ 
pattern corresponding to a $N=5$ array with various junction spacings and sizes. 
This geometry can be realized by optical lithography. Other patterns where 
$I_{\rm max}(\tilde{H})$ is zero on a interval can also be generated.

While at LERMA, M.-H. Chung of Taeduk Observatory developed the code for $SIS$ 
mixing in $DJTLs$. We acknowledge Y. Delorme, F. Dauplay, B. Lecomte, A. Feret 
and J.-M. Krieg for their contribution to the $SIS$ mixer work.

\end{document}